\documentclass[preprint,showpacs,preprintnumbers,amsmath,amssymb,aps,pre]{revtex4}

\usepackage[utf8]{inputenc} 
\usepackage{epsfig}
\usepackage{natbib}
\usepackage{amsmath}
\usepackage{subfigure}
\usepackage{epstopdf} 
\usepackage{graphicx}
\usepackage{color}
\usepackage{pict2e}
\newcommand{\comment}[1]{}

\begin{document}

\title{Transient chaos under coordinate transformations in relativistic systems}

\author{D. S. Fern\'{a}ndez}

\author{Á. G. L\'{o}pez}

\author{J. M. Seoane}

\author{M. A. F. Sanju\'{a}n}
\affiliation{Nonlinear Dynamics, Chaos and Complex Systems Group, Departamento de
F\'{i}sica, Universidad Rey Juan Carlos, Tulip\'{a}n s/n, 28933 M\'{o}stoles, Madrid, Spain}

\date{\today}

\begin{abstract}
We use the H\'{e}non-Heiles system as a paradigmatic model for chaotic scattering to study the Lorentz factor effects on its transient chaotic dynamics. In particular, we focus on how time dilation occurs within the scattering region by measuring the time in a clock attached to the particle. We observe that the several events of time dilation that the particle undergoes exhibit sensitivity to initial conditions. However, the structure of the singularities appearing in the escape time function remains invariant under coordinate transformations. This occurs because the singularities are closely related to the chaotic saddle. We then demonstrate using a Cantor-like set approach that the fractal dimension of the escape time function is relativistic invariant. In order to verify this result, we compute by means of the uncertainty dimension algorithm the fractal dimensions of the escape time functions as measured with inertial and comoving with the particle frames. We conclude that, from a mathematical point of view, chaotic transient phenomena are equally predictable in any reference frame and that transient chaos is coordinate invariant.
\end{abstract}

\pacs{05.45.Ac,05.45.Df,05.45.Pq}
\maketitle

\section{Introduction} \label{sec:1}

Chaotic scattering in open Hamiltonian systems is a fundamental part of the theoretical study of dynamical systems. There are many applications such as the interaction between the solar wind and the magnetosphere tail \cite{seoane2013}, the simulation in several dimensions of the molecular dynamics \cite{lin2013}, the modeling of chaotic advection of particles in fluid mechanics \cite{daitche2014}, or the analysis of the escaping mechanism from a star cluster or a galaxy \cite{zotos2017,navarro2019}, to name a few. A scattering phenomenon is a process in which a particle travels freely from a remote region and encounters an obstacle, often described in terms of a potential, which affects its evolution. Finally, the particle leaves the interaction region and continues its journey freely. This interaction is typically nonlinear, possibly leading the particle to perform transient chaotic dynamics, i.e., chaotic dynamics with a finite lifetime \cite{lai2010,grebogi1983}. Scattering processes are commonly studied by means of the scattering functions, which relate the particle states at the beginning of its evolution once the interaction with the potential has already taken place. Thus, nonlinear interactions can make these functions exhibit self-similar arrangements of singularities, which hinder the system predictability \cite{aguirre2009}. Transient chaos is a manifestation of the presence in phase space of a chaotic set called non-attracting chaotic set, also called chaotic saddle \cite{ott1993}. This phenomenon can be found in a wide variety of situations \cite{tel2015}, as for example the dynamics of decision making, the doubly transient chaos of undriven autonomous mechanical systems or even in the sedimentation of volcanic ash.

There have been numerous efforts to characterize chaos in relativistic systems in an observer-independent manner \cite{hobill1994}. It has been rigorously demonstrated that the sign of the Lyapunov exponents is invariant under coordinate transformations that satisfy four minimal conditions \cite{motter2003}. More specifically, such conditions consider that a valid coordinate transformation has to leave the system \textit{autonomous}, its phase space \textit{bounded}, the invariant measure \textit{normalizable} and the domain of the new time parameter \textit{infinite} \cite{motter2003}. As a consequence, chaos is a property of relativistic systems independent of the choice of the coordinate system in which they are described. In other words, homoclinic and heteroclinic tangles cannot be untangled by means of coordinate transformations. We shall utilize the Lorentz transformations along this paper, which satisfy this set of conditions \cite{motter2009}. Although we utilize a Hamiltonian system in its open regime, from the point of view of Lyapunov exponents the phase space can be considered bounded because of the presence of the chaotic saddle. This set is located in a finite region of the system's phase space and contains all the non-escaping orbits in the hyperbolic regime. Hence, the Lyapunov exponents are well-defined because these trajectories stay in the saddle forever. On the other hand, concerning the computation of the escape time function, we shall only consider along this work the finite part of the phase space where the escaping orbits remain bounded, and similarly from the point of view of the finite-time Lyapunov exponents the phase space can be considered bounded as well \cite{vallejo2003}.

Despite the fact that the sign of the Lyapunov exponents is invariant, the precise values of these exponents,  which indicate ``how chaotic" a dynamical system is, are noninvariant. Therefore, this lack of invariance leaves some room to explore how coordinate transformations affect the unpredictability in dynamical systems with transient chaos. In the present work we analyze the structure of singularities of the scattering functions under a valid coordinate transformation. In particular, we compute the fractal dimension of the escape time function as measured in an inertial reference frame and another non-inertial reference frame comoving with the particle, respectively. We then characterize the system unpredictability by calculating this fractal dimension, since it enables to infer the dimension of the chaotic saddle \cite{aguirre2001}. Indeed, this purely geometrical method has been proposed as an independent-observer procedure to determine whether the system behaves chaotically \cite{motter2001}.

Relevant works have been devoted to analyze the relationship between relativity and chaos in recent decades
\cite{barrow1982,chernikov1989,ni2012}. More recently, the Lorentz
factor effects on the dynamical properties of the system have also
been studied in relativistic chaotic scattering
\cite{bernal2017,bernal2018}. In this paper, we focus on how changes
of the reference frame affect typical phenomena of chaotic
scattering. We describe the model in Sec.~\ref{sec:2}, which
consists of a relativistic version of the H\'{e}non-Heiles system.
Two well-known scattering functions are explored in
Sec.~\ref{sec:3}, such as the exit through which the particle
escapes and its escape time. In Sec.~\ref{sec:4}, we demonstrate the
fractal dimension invariance under a coordinate transformation by
using a Cantor-like set approach. Subsequently, we quantify the
unpredictability of the escape times and analyze the effect of such
a reference frame modification. We conclude with a discussion of the
main results and findings of the present work in Sec.~\ref{sec:5}.

\section{Model description} \label{sec:2}

\begin{figure}[b!]
	\centering
	\includegraphics[width=0.35\textwidth]{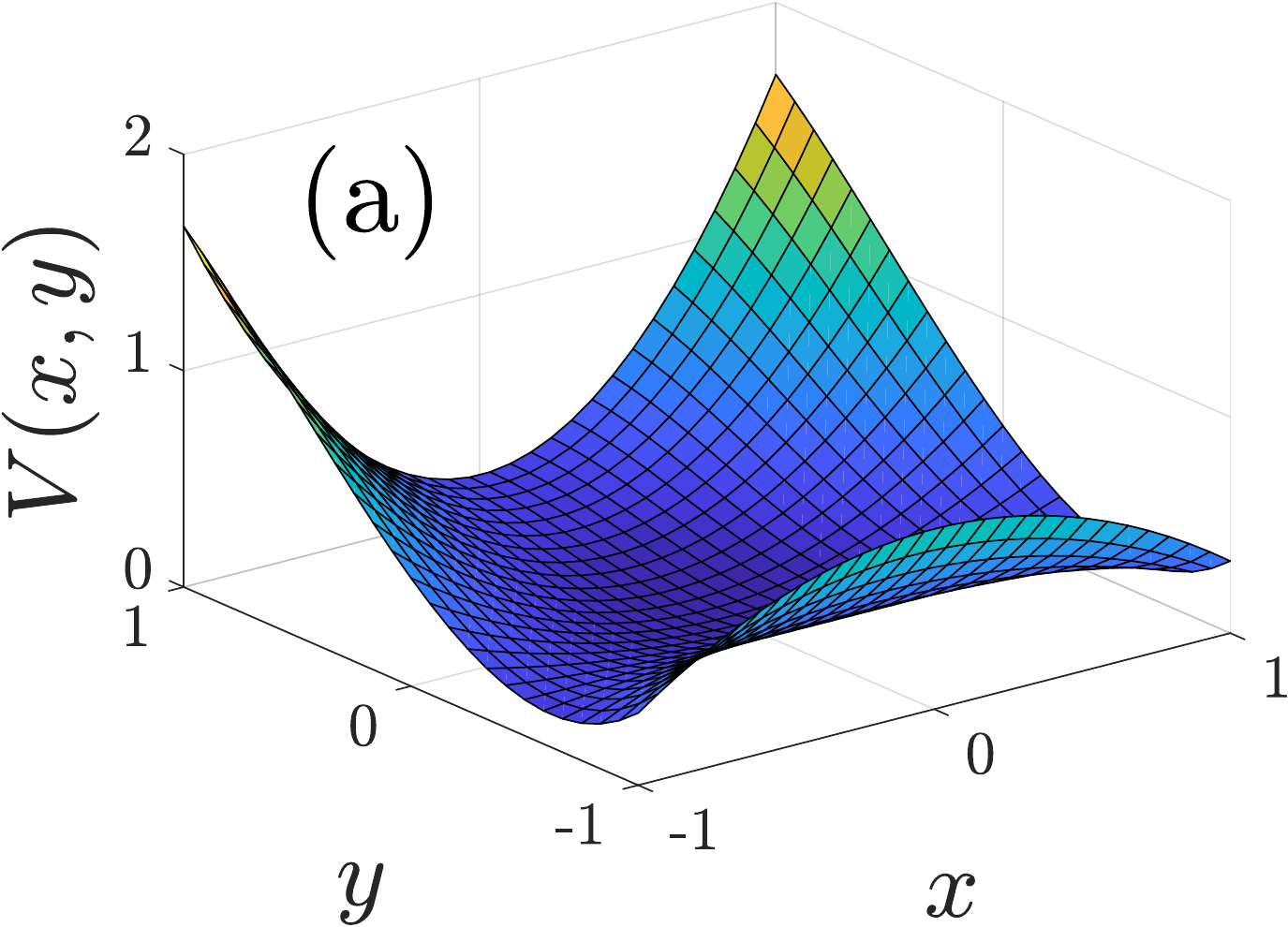}
	\includegraphics[width=0.35\textwidth]{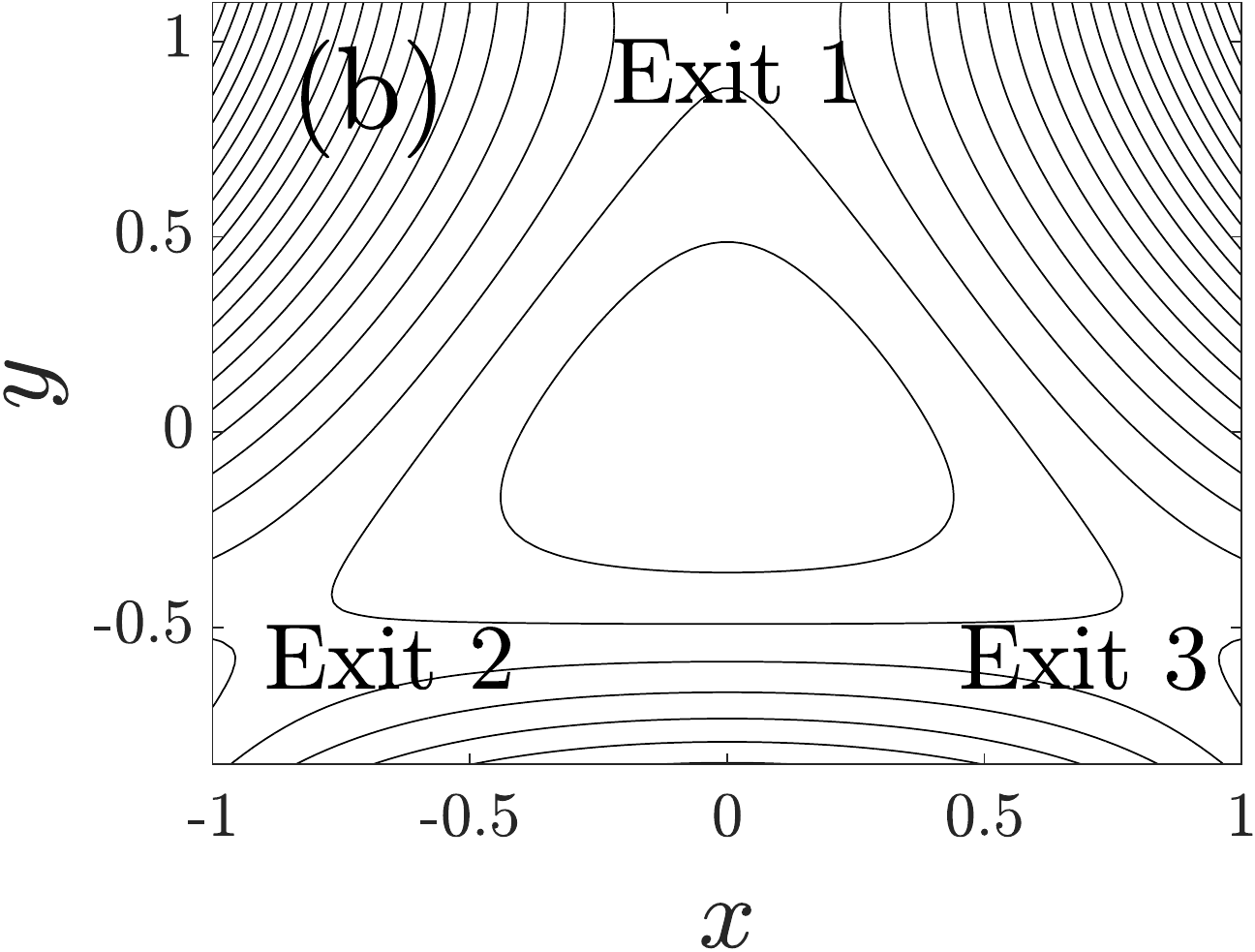}
	\caption{(a) The three-dimensional representation of the H\'{e}non-Heiles potential $V(x,y) = \frac{1}{2}(x^2 + y^2) + x^2 y - \frac{1}{3}y^3$. (b) The isopotential curves in the physical space show that the H\'{e}non-Heiles system is open and has triangular symmetry. If the energy of the particle is higher than a threshold value, related to the potential saddle points, there exist unbounded orbits. Following these trajectories the particle leaves the scattering region through any of the three exits.}
	\label{fig:sec2.1}
\end{figure}

The H\'{e}non-Heiles system was proposed in 1964 to study the existence of a third integral of motion in galactic models with axial symmetry \cite{henon1964}. We consider a single particle whose total mechanical energy can be denoted as $E_N$ in the Newtonian approximation. This energy is conserved along the trajectory described by the particle, which is launched from the interior of the potential well, within a finite region of the phase space called the scattering region. We have utilized a dimensionless form of the H\'{e}non-Heiles system, so that the potential is written as
\begin{equation} V(x,y) = \frac{1}{2}(x^2 + y^2) + x^2 y - \frac{1}{3}y^3, \label{eq:V}
\end{equation}
where $x$ and $y$ are the spatial coordinates. When the energy is above a threshold value, the potential well exhibits three exits due to its triangular symmetry in the physical space, i.e., the plane $(x,y)$, as visualized in Fig.~\ref{fig:sec2.1}. We call Exit 1 the exit located at the top $(y \to +\infty)$, Exit 2 the one located downwards to the left $(x \to - \infty, y \to -\infty)$ and Exit 3 the one at the right $(x \to + \infty, y \to -\infty)$. One of the characteristics of open Hamiltonian systems with escapes is the existence of highly unstable periodic orbits known as Lyapunov orbits \cite{contopoulos1990}, which are placed near the saddle points. In fact, when a trajectory crosses through a Lyapunov orbit, it escapes to infinity and never returns back to the scattering region. Furthermore, we recall that the energy of the particle determines also the dynamical regime. We can distinguish two open regimes in which escapes are allowed. On the one hand, in the nonhyperbolic regime the KAM tori coexist with the chaotic saddle and the phase space exhibits regions where dynamics is regular and also chaotic \cite{sideris2006}, whereas the chaotic saddle rules the dynamics in the hyperbolic regime, making it completely chaotic.

When the speed of the particle is comparable to the speed of light, the relativistic effects have to be taken into account \cite{ohanian2001}. In the present work we consider a particle which interacts in the limit of weak external fields, and therefore we deal with a special relativistic version of the H\'{e}non-Heiles system, whose dynamics is governed by the conservative Hamiltonian \cite{lan2011,chanda2018,kovacs2011,calura1997} \begin{equation} H = c \sqrt{c^2 + p^2 + q^2} + V(x,y), \label{eq:H} \end{equation} where $c$ is the value of the speed of light, and $p$ and $q$ are the momentum coordinates. On the other hand, the Lorentz factor is defined as \begin{equation} \gamma = \frac{1}{\sqrt{1 - \frac{\textbf{v}^2}{c^2}}} = \frac{1}{\sqrt{1 - \beta^2}}, \label{eq:gamma} \end{equation} where $\textbf{v}$ is the velocity vector of the particle and $\beta = |\textbf{v}|/c$ the ratio between the speed of the particle and the speed of light. The Lorentz factor $\gamma$ and $\beta$ are two equivalent ways to express how large is the speed of the particle compared to the speed of light. These two factors vary in the ranges $\gamma \in [1,+\infty)$ and $\beta \in [0,1)$, respectively. For convenience, we shall use $\beta$ as a parameter along this work. Hamilton's canonical equations can be derived from Eq.~\eqref{eq:H}, yielding the equations of motion \begin{equation} \begin{aligned} \dot{x} = & \frac{\partial H}{\partial p} = \frac{p}{\gamma}, & \dot{p} = -\frac{\partial H}{\partial x} =-x-2xy, \\ \dot{y} = & \frac{\partial H}{\partial q} = \frac{q}{\gamma}, & \dot{q} = -\frac{\partial H}{\partial y} = y^2 -x^2 -y, \end{aligned} \label{eq:odes} \end{equation} where the Lorentz factor can be alternatively written in the momentum-dependent form as $\gamma = \frac{1}{c} \sqrt{c^2 + p^2 + q^2}$.  Although the complete phase space is four-dimensional, the conservative Hamiltonian constrains the dynamics to a three-dimensional manifold of the phase space, known as the energy shell.

Some recent works aim at isolating the effects of the variation of the Lorentz factor $\gamma$ (or $\beta$ equivalently) from the remaining variables of the system \cite{bernal2017,bernal2018}. In order to accomplish this, they modify the initial value of $\beta$ and use it as the only parameter of the dynamical system. Since $\beta$ is a quantity that depends on $|\textbf{v}|$ and $c$, they choose to vary the numerical value of $c$. Needless to say, the value of the speed of light $c$ remains constant during the particle trajectory. The fundamental reason for deciding to increase the kinetic energy of the system by reducing the numerical value of the speed of light is simply as follows. If we keep the H\'{e}non-Heiles potential constant and increase the speed of the particle to values close to the speed of light, the potential will be in a much lower energy regime compared to the kinetic energy of the particle. Therefore, the potential becomes negligible and the interaction between them becomes irrelevant. Consequently, each time we select a value of the speed of light we are scaling the system, and hence the ratio of the kinetic energy and the potential as well. The sequence of potential wells with different values of $\beta$ represents potential wells with the H\'{e}non-Heiles morphology, but at different scales in which the interaction of a relativistic particle is not trivial. In this way, the effects of the Lorentz factor on the dynamics are isolated from the other system variables, because the Lorentz factor is the only parameter that differentiates all these scaled systems.

\begin{figure}[htp!]
	\centering
	\includegraphics[width=0.31\textwidth]{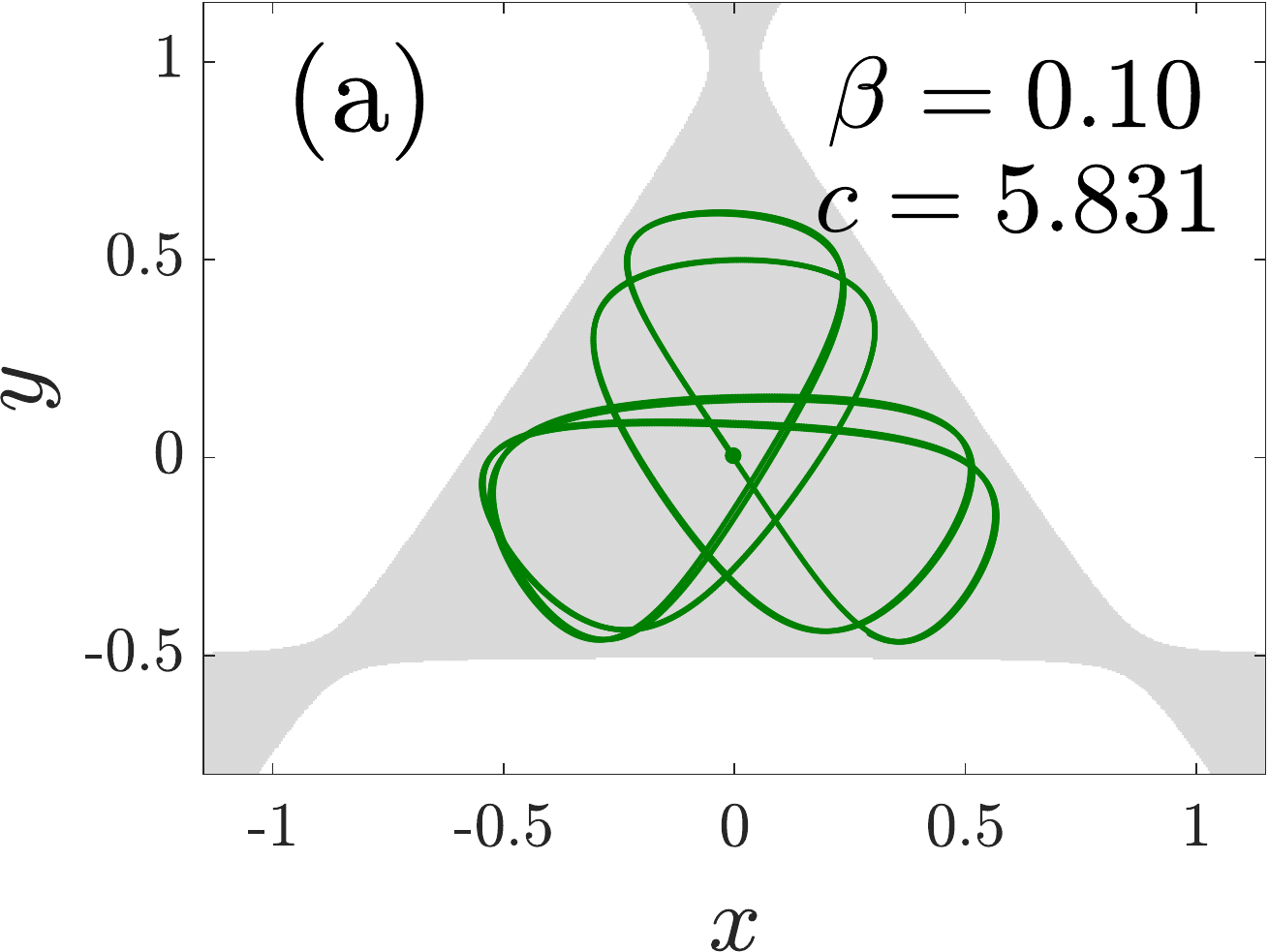}
	\quad
	\includegraphics[width=0.31\textwidth]{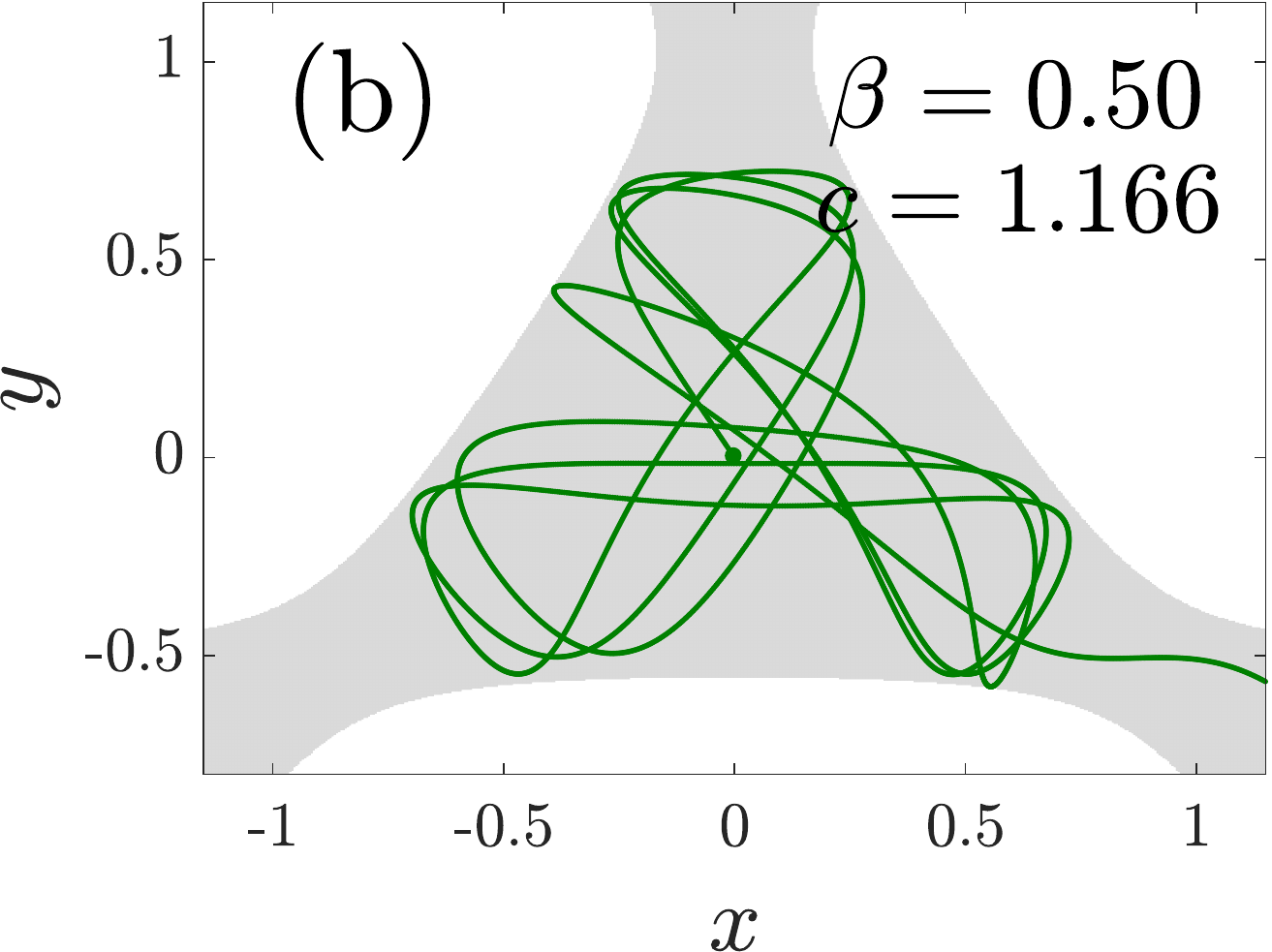}
	\quad
	\includegraphics[width=0.31\textwidth]{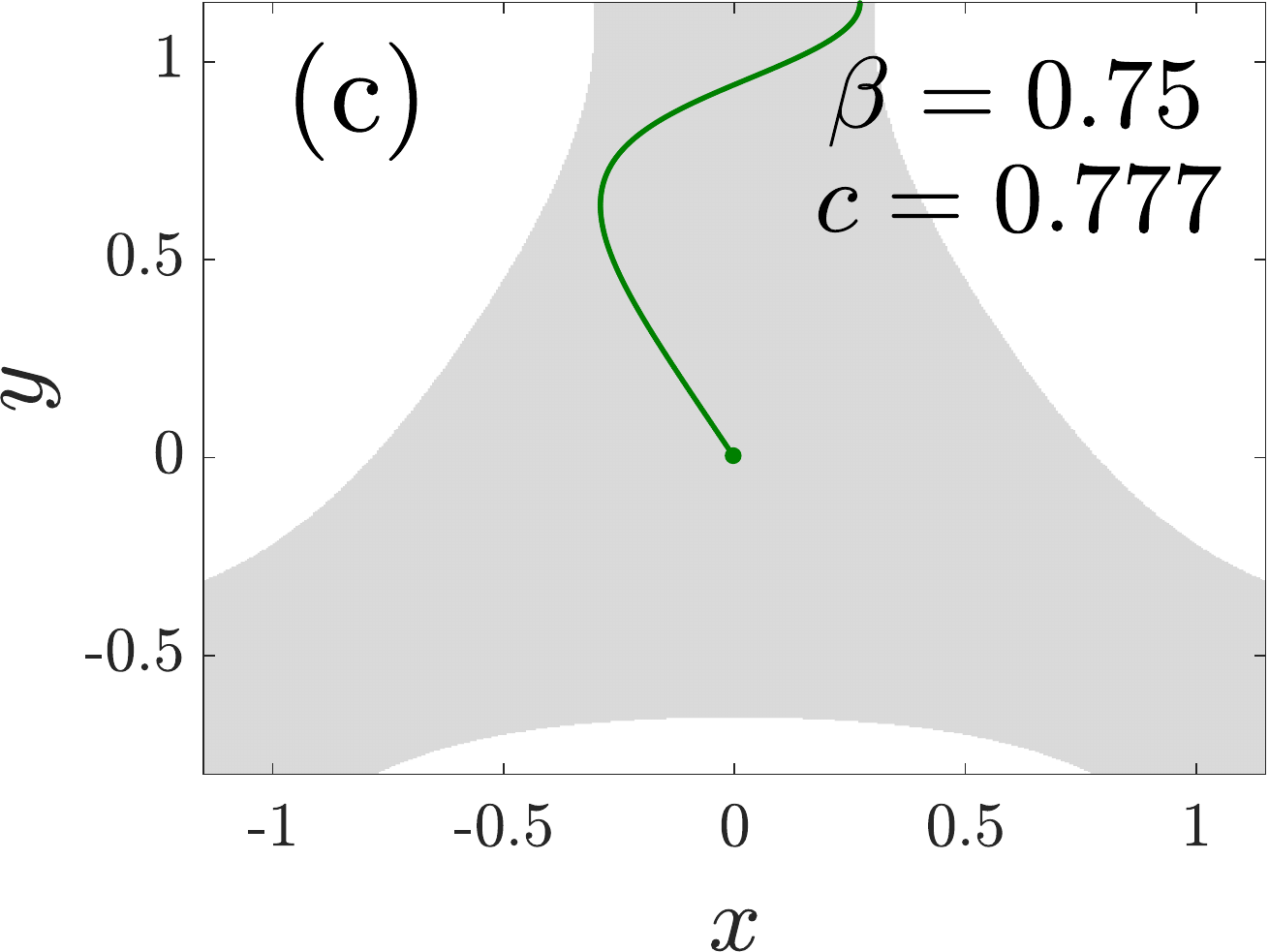}
	\caption{The evolution of a particle launched within the scattering region from the same initial condition for different values of $\beta$. (a) For a very low $\beta$ (Newtonian approximation), the particle is trapped in the KAM tori and describes a bounded trajectory. (b) The value of $\beta$ is large enough to destroy the KAM tori and the particle leaves the scattering region following a trajectory typical of transient chaos. (c) Finally, a larger value of $\beta$ than in (b) makes the particle escape faster.}
	\label{fig:sec2.2}
\end{figure}

We then consider the same initial value of the particle speed $|\textbf{v}_0|$ in every simulation with a different value of $\beta$, launching the particle from the minimum potential, which is located at $(x_0,y_0) = (0,0)$ and where the potential energy is null. We have arbitrarily chosen $|\textbf{v}_0| \approx 0.5831$ (as in \cite{bernal2017,bernal2018}), which corresponds to the open nonhyperbolic regime with energy $E_N = 0.17$, close to the escape energy in the Newtonian approximation. Thus, we analyze how the relativistic parameter $\beta$, as its value increases, affects the dynamical properties starting from the nonhyperbolic regime. The numerical value of $c$ varies, as shown in Fig.~\ref{fig:sec2.2}, and for instance if the simulation is carried out for a small $\beta$, where $|\textbf{v}_0| \ll c$, the initial speed of the particle only represents a very low percentage of the speed of light. In this case, we recover the Newtonian approximation and the classical version of the H\'{e}non-Heiles system. On the contrary, if the simulation takes place with a value of $\beta$ near one, the speed of the particle represents a high percentage of the speed of light and the relativistic effects on the dynamics become more intense.

Numerical computations reveal that the KAM tori are mostly destroyed at $\beta \approx 0.4$, and hence the dynamics is hyperbolic for higher values of $\beta$ \cite{bernal2018}. If some small tori survive, they certainly do not rule the system overall dynamics. As we focus on the hyperbolic regime, the simulations are run for values of $\beta \in [ 0.5, 0.99]$ and by means of a fixed step fourth-order Runge-Kutta method \cite{press1992}. We recall that the initial values of the momentum $(p_0,q_0)$ depend on the chosen initial value of $\beta$, and therefore this computational technique (to vary the value of $\beta$ fixing $|\textbf{v}_0|$) is an ideal method to increase the particle kinetic energy to the relativistic regime. For example, a particle trapped in the KAM tori can escape if the initial value of $\beta$ is high enough, as shown in Fig.~\ref{fig:sec2.2}.

\section{Escape times in inertial and non-inertial frames} \label{sec:3}

\begin{figure}[b!]
	\centering
	\raisebox{.11\height}{\includegraphics[width=0.385\textwidth]{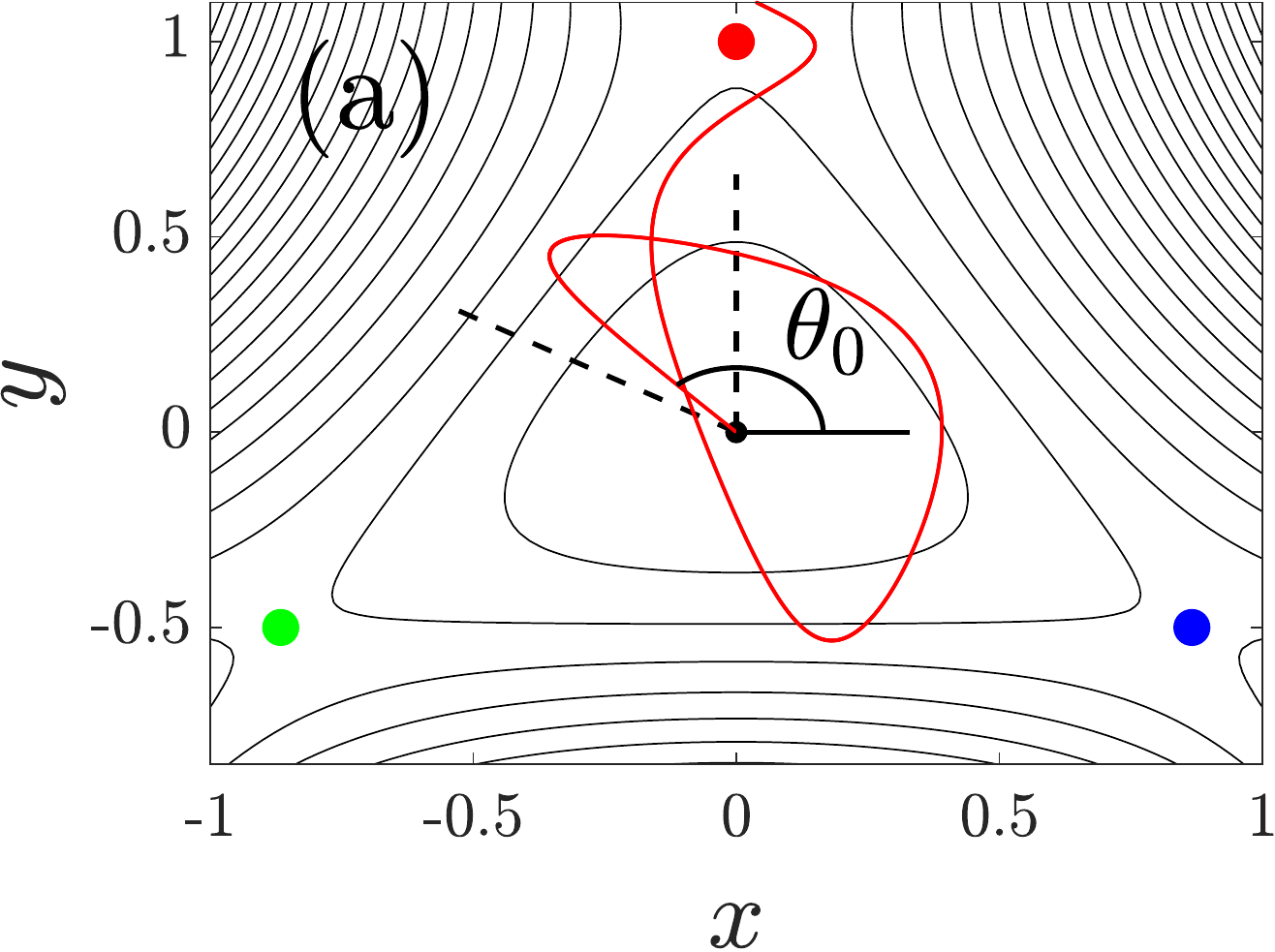}}
	\quad
	\includegraphics[width=0.475\textwidth]{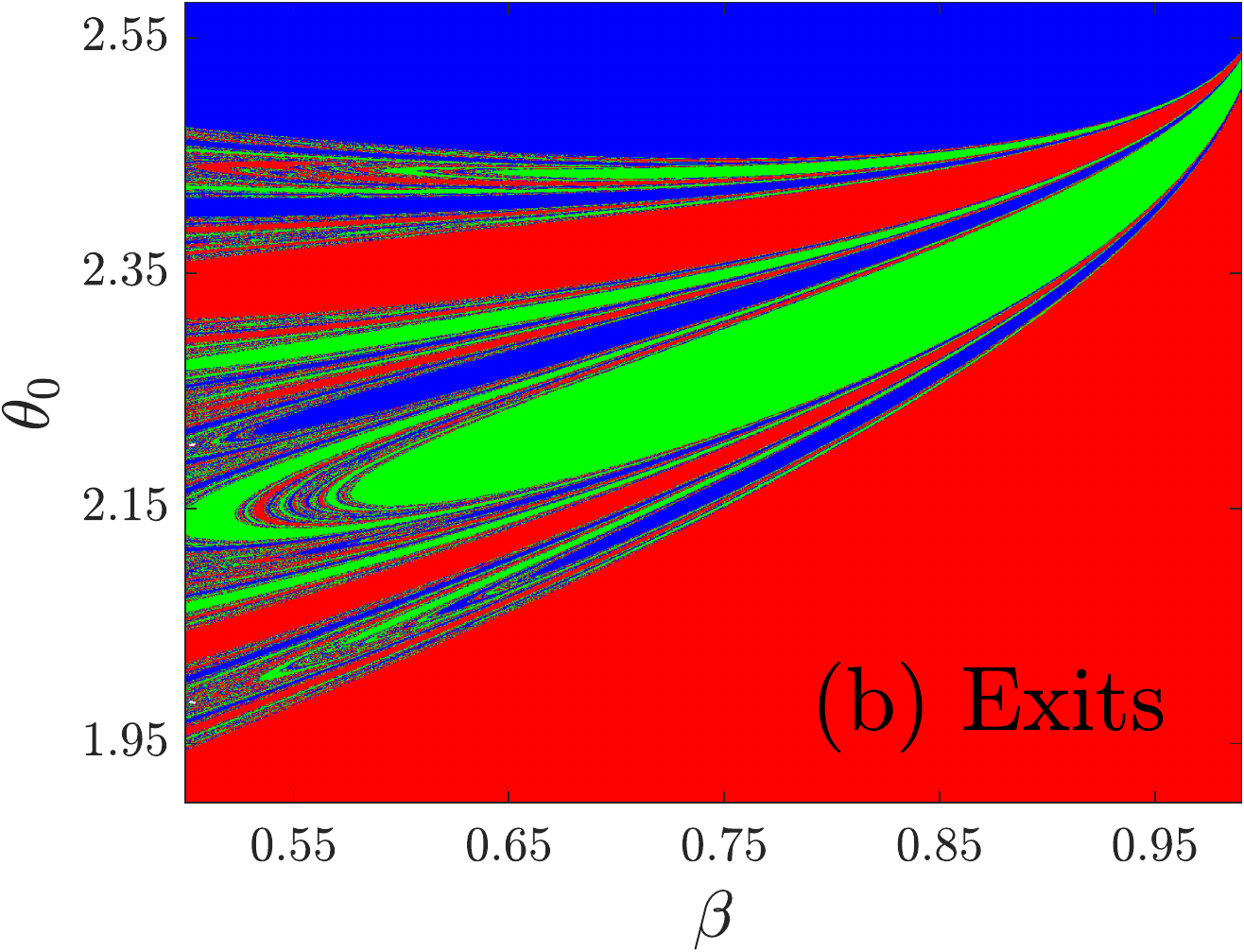}\\
	\caption{(a) Each of the exits is identified with a different color, such that Exit 1 (red), Exit 2 (green) and finally Exit 3 (blue). In order to avoid redundant results due to the triangular symmetry of the well, we only let the particle evolve from the angular region $\theta_0 \in \left[\pi/2, 5\pi/6 \right]$ (black dashed lines). (b) The scattering function of the exits $(2000 \times 2000)$ given the parameter map $(\beta \in [0.5,0.99],\theta_0 \in [\pi/2, 5\pi/6])$ in the hyperbolic regime.}
	\label{fig:sec3.1}
\end{figure}

The scattering functions enable us to represent the relation between input and output dynamical states of the particle, i.e., how the interaction of the particle with the potential takes place. The potential of H\'{e}non-Heiles leads the particle to describe chaotic trajectories before converging to a specific exit, which makes the scattering functions exhibit a fractal structure. In order to verify the sensitivity of the system to exits and escape times, we launch particles from the potential minimum slightly varying the shooting angle $\theta$ that is formed by the initial velocity vector and the positive $x$-axis, as shown in Fig.~\ref{fig:sec3.1}(a).

The maximum value of the kinetic energy is reached at the potential minimum, as the system is conservative. We define the value of the Lorentz factor associated with this maximum kinetic energy as the critical Lorentz factor \begin{equation} \gamma_c (\beta) = \frac{1}{\sqrt{1-\beta^2}}.\end{equation} We emphasize that the initial Lorentz factor of every particle is the critical Lorentz factor, since every trajectory is initialized from the potential minimum in this work. We shall monitor the Lorentz factor of the particle along its trajectory and use the critical Lorentz factor as the criterion of whether the particle has escaped or not. This escape criterion is based on the fact that the value of the kinetic energy remains bounded while the particle evolves chaotically within the potential well, bouncing back and forth against the potential barriers before escaping. The Lorentz factor value then varies between the unity and the critical value inside the scattering region, i.e., $\gamma(t) \in [1, \gamma_c]$. Eventually, the particle leaves the scattering region and the value of its Lorentz factor breaks out towards infinity, because its kinetic energy does not remain bounded anymore. In order to prevent this asymptotic behavior of the Lorentz factor, it is convenient to set that the escape happens at the time $t_e$ when $\gamma(t_e)>\gamma_c$. In this manner, we define the scattering region as the part of the physical space where the dynamics is bounded. This escape criterion is computationally affordable and useful to implement in any Hamiltonian system without knowing specific information about the exits. In addition, it includes all the escapes that take place when the Lyapunov orbit criterion is considered.

A particle launched with $\theta = \pi/2$ escapes directly towards the Exit 1 for every value of $\beta$ as shown in Fig.~\ref{fig:sec3.1}(b), whereas if it is launched with $\theta = 5\pi/6$ the particle bounces against the potential barrier placed between Exit 1 and Exit 2 and escapes through the Exit 3. The whole structure of exits in between is apparently fractal. Nonetheless, the exit function becomes smoother when the value of $\beta$ increases, but it is never completely smooth. On the other hand, we recall that the chaotic saddle is an observer-independent set of points formed by the intersection of the stable and unstable manifolds. Concretely, the stable manifold of an open Hamiltonian system is defined as the boundary between the exit basins \cite{ott1993}. If a particle starts from a point arbitrarily close to the stable manifold it will spend an infinite time in converging to an exit, i.e., it never escapes. The unstable manifold is the set along which particles lying infinitesimally close to the chaotic saddle will eventually leave the scattering region in the course of time \cite{tel2015}.

The escape time can be easily defined as the time the particle spends evolving inside the scattering region before escaping to infinity. In nonrelativistic systems, the particular clock in which the time is measured is irrelevant since time is absolute. However, here we consider two time quantities: the time $t$ that is measured by an inertial reference frame at rest and the \textit{proper time} $\tau$ as measured by a non-inertial reference frame comoving with the particle. This proper time is simply the time measured by a clock attached to the particle.

As is well known, an uniformly moving clock runs slower by a factor $\sqrt{1-\beta^2}$ in comparison to another identically constructed and synchronized clock at rest in an inertial frame. Therefore, we assume that at any instant of time the clock of the accelerating particle advances at the same rate as an inertial clock that momentarily had the same velocity \cite{barton1999}. In this manner, given an infinitesimal time interval $dt$, the particle clock will measure a time interval \begin{equation}
d\tau = \frac{dt}{\gamma(t)},
\label{eq:dtau}
\end{equation} where $\gamma(t)$ is the particle Lorentz factor at the instant of time $t$. Since the Lorentz factor is greater than the unity, the proper time interval always obeys that $d\tau \le dt$, which is just the mathematical statement of the twin paradox. When the particle velocities are very close to the speed of light, the time dilation phenomenon takes place so that the time of the particle clock runs more slowly in comparison to clocks at rest in the potential. In the context of special relativity, it is important to bear in mind that it is assumed that the potential does not affect the clocks rate. In other words, all the clocks placed at rest in any point of the potential are ticking at the same rate along this work.

\begin{figure}[b!]
	\centering
	\includegraphics[width=0.48\textwidth]{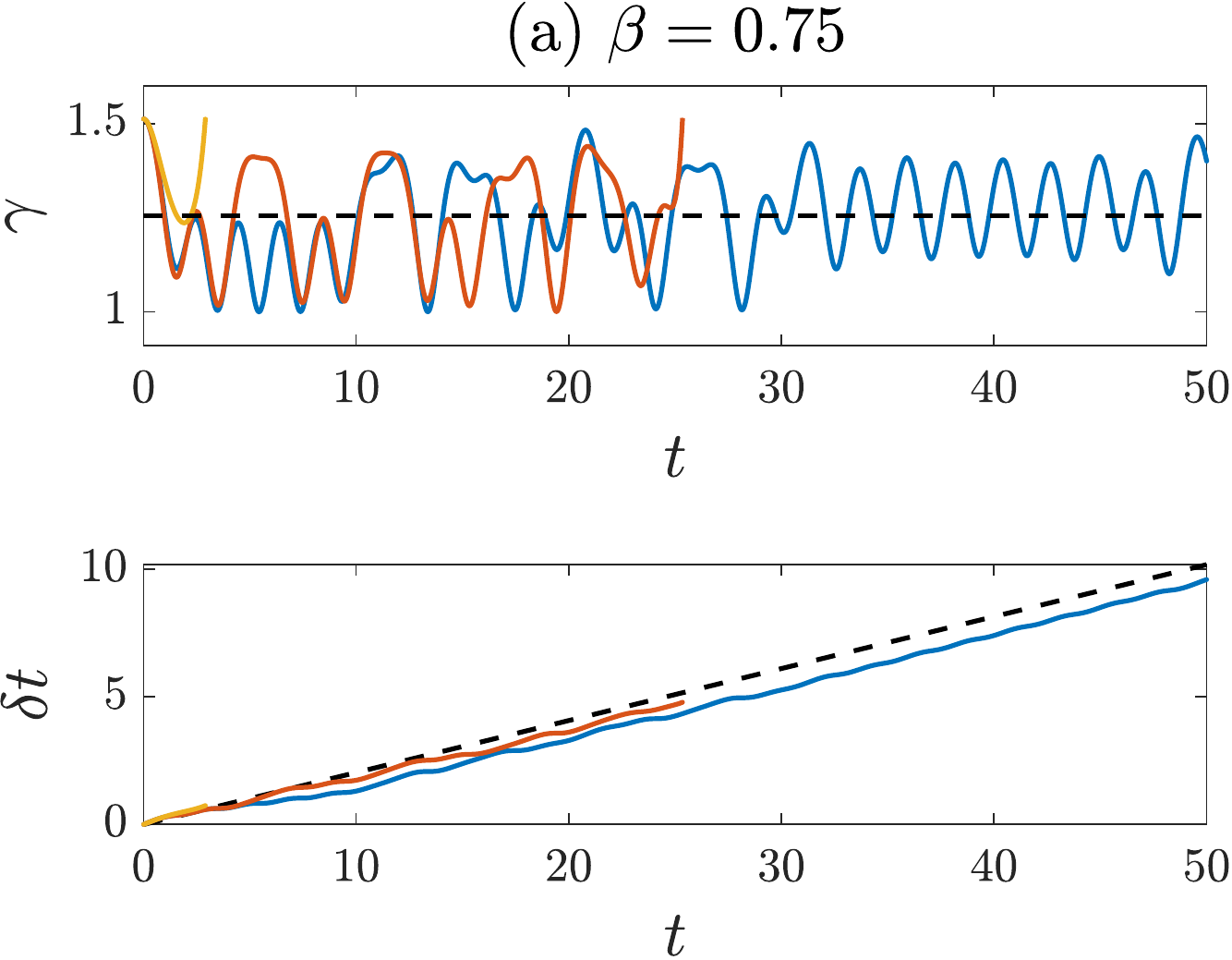}
	\quad
	\includegraphics[width=0.475\textwidth]{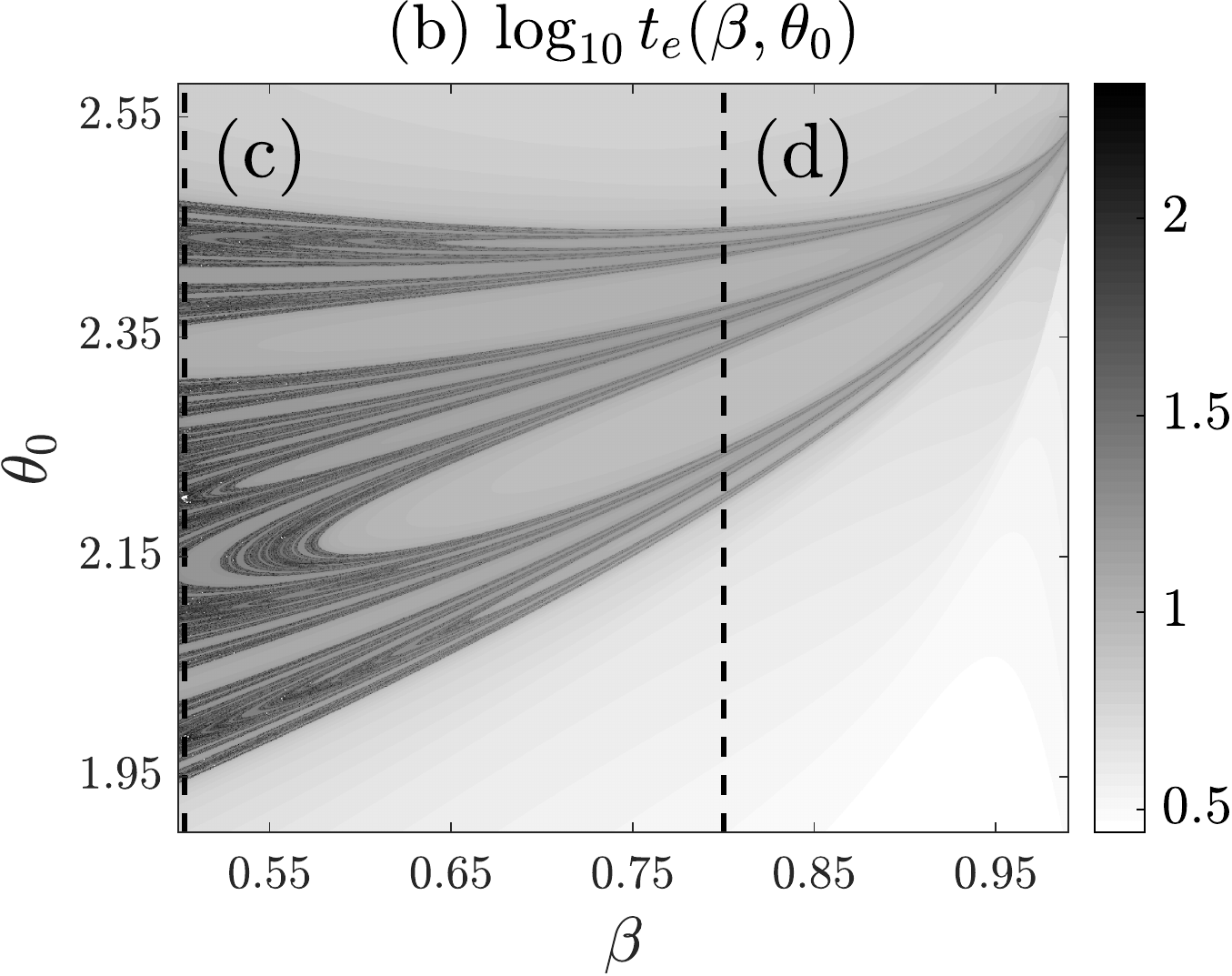} \\
	\bigskip
	\includegraphics[width=0.475\textwidth]{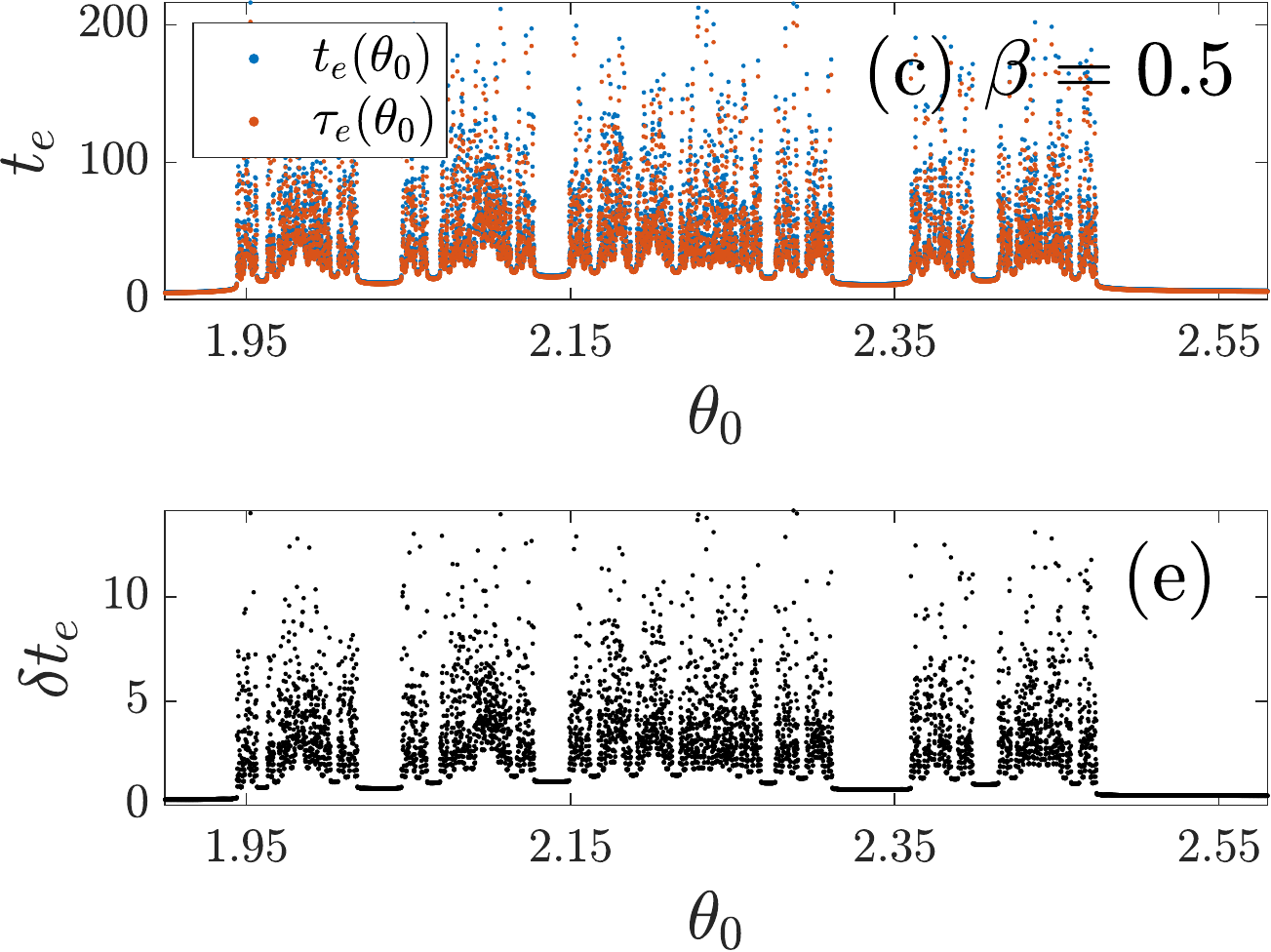}
	\quad
	\includegraphics[width=0.47\textwidth]{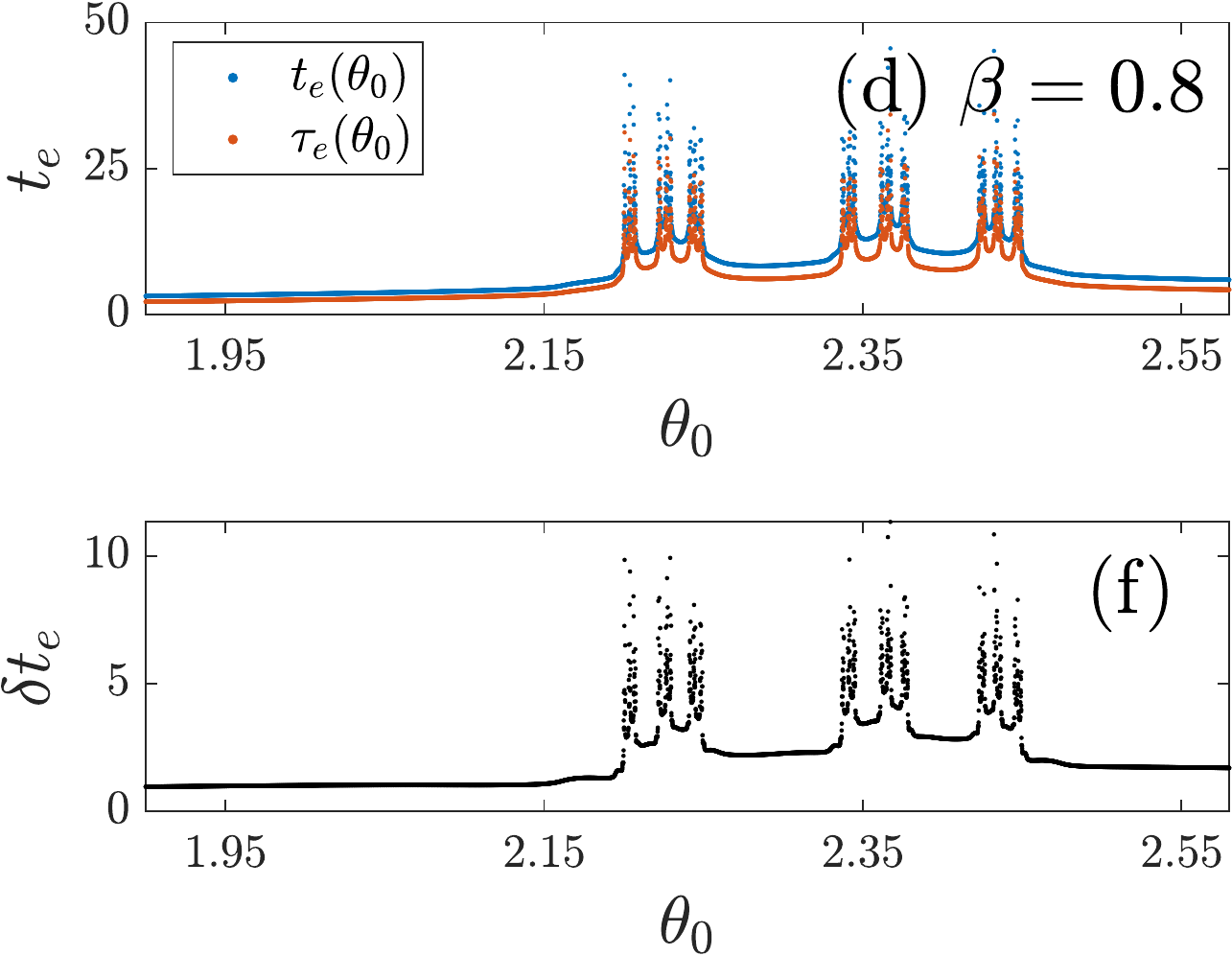}
	\caption{(a) The Lorentz factor evolution $\gamma (t)$ of three different trajectories: a fast escape (yellow) and two typical transient chaotic trajectories (red and blue). The dashed guideline represents the Lorentz factor value of $\bar{\gamma}$ (black), corresponding to $\beta = 0.75$. The time differences $\delta t (t)$ along these trajectories is also shown. (b) The scattering function of escape times $t_e$ in logarithmic scale given the parameter map $(\beta \in [0.5,0.99],\theta_0 \in [\pi/2, 5\pi/6])$. The two black dashed lines corresponds to the subfigures (c) and (d), which show the scattering function of escape time $t_e(\theta_0)$ (blue) and $\tau_e(\theta_0)$ (red) for $\beta = 0.5$ and $\beta = 0.8$, respectively. (e, f) The time difference function $\delta t_e (\theta_0)$ (black) for the same values of $\beta$.}
	\label{fig:sec3.2}
\end{figure}

Without loss of generality, Eq.~\ref{eq:dtau} can be expressed as an integral in the form \begin{equation}
\tau_e = \int_{0}^{t_e} \frac{dt}{\gamma(t)},
\label{eq:tau_e}
\end{equation} where the final time of the integration interval is the escape time in the inertial frame. We shall solve this integral using the Simpson's rule \cite{jeffreys1988}. Since each evolution of the Lorentz factor is unique because each particle describes a distinct chaotic trajectory, every particle clock measures a different proper time at any instant of time $t$. Nonetheless, as the dynamics is bounded in the same energetic conditions given a value of $\beta$, the Lorentz factor of all trajectories is similar on average at any instant of time $t$. For this reason, we assume that there exists an average value of the Lorentz factor along the particle trajectory, and estimate it as the arithmetic mean between the maximum and minimum values of the bounded Lorentz factor inside the scattering region, i.e., \begin{equation}
\bar{\gamma}(\beta) = \frac{1+\gamma_c}{2} = \frac{1+\sqrt{1-\beta^2}}{2 \sqrt{1-\beta^2}}.
\label{eq:average_gamma}
\end{equation}
Using this definition to Eq.~\ref{eq:tau_e}, we can define an average time dilation in the form $\bar{\tau}_e \equiv t_e/\bar{\gamma}$. This value should only be regarded as an approximation, which shall prove of great usefulness to interpret the numerical results obtained ahead. Accordingly, the difference between both the average escape time and the time $t_e$ is also approximately linear on average. In this manner, we can also define the magnitude
\begin{equation}
\delta \bar{t}_e \equiv t_e - \bar{\tau}_e = \frac{1-\sqrt{1-\beta^2}}{1+\sqrt{1-\beta^2}}t_e.
\label{eq:tau_e3}
\end{equation} We emphasize that this value is again just an approximation representing the average behavior of the system, which disregards the fluctuations of the Lorentz factor. It reproduces qualitatively the behavior when the dynamics is bounded in the well, as shown in Fig.~\ref{fig:sec3.2}(a).

The escape time function is similar to the exit function, as shown in Fig.~\ref{fig:sec3.2}(b); the longest escape times are located close to the the boundary of the exit regions, i.e., the mentioned stable manifold, because these trajectories spend long transient times before escaping. In this manner, the structure of singularities is again associated to the stable manifold, equally that the exit function. This is an evidence that the fractality of the escape time function must be an observer-independent feature, since the exit through which the particle escapes does not depend on the considered clock. Indeed, we observe that the escape proper time function exhibits a similar structure of singularities because of the approximated linear relation described by $\bar{\tau}_e$ (see Figs.~\ref{fig:sec3.2}(c) and \ref{fig:sec3.2}(d)). Despite being almost identical structures, the dilation time phenomenon always makes $\tau_e(\theta_0) < t_e(\theta_0)$.

Importantly, the time difference function $\delta t_e(\theta_0)$ also preserves the fractal structure as shown in Figs.~\ref{fig:sec3.2}(e) and \ref{fig:sec3.2}(f). This occurs because sensitivity to initial conditions is translated into sensitivity to time dilation phenomena. The longer the time the particle spends in the well, the more travels from the center to the potential barriers and back. If we think of each of these travels as an example of a twin paradox journey, we get an increasing time dilation for particles that spend more time in the well. Since these times are sensitive to modifications in the initial conditions, so are time dilation effects. We could then introduce what might be called the triplet paradox. In this case an additional third sibling leaves the planet and comes back to the starting point having a different age than their two other siblings, because of the sensitivity to initial conditions. This phenomenon in particular illustrates how chaotic dynamics affects typical relativistic phenomena.

\section{Invariant fractal dimension and persistence of transient chaos} \label{sec:4}

The chaotic saddle and the stable manifold are self-similar fractal sets when the underlying dynamics is hyperbolic \cite{ott1993}. This fact is reflected in the peaks structure of the escape time functions, which is present at any scale of initial conditions. In this sense, the escape time functions share with the Cantor set some properties with regard to their singularities, and therefore to their fractal dimensions. It is possible to study the fractal dimensions of the escape time functions in terms of a Cantor-like set \cite{lau1991,seoane2007}.

In this manner, we can build a Cantor-like set to schematically represent the escape of particles launched from different initial conditions $\theta_0$. We consider that a certain fraction $\eta_t$ of particles escapes from the scattering region when a minimal characteristic time $t_0$ has elapsed. If these particles were launched from initial conditions centered in the original interval, two identical segments are created; the trajectories that began in those segments do not escape at least by a time $t_0$. Similarly, a same fraction of particles $\eta_t$ from the two surviving segments escapes by a time $2t_0$. If we continue this iterative procedure for $3t_0$, $4t_0$ and so on, we obtain a Cantor-like set of Lebesgue measure zero with associated fractal dimension $d_t$ that can be computed as \begin{equation}
d_t = \frac{\ln 2}{\ln 2 - \ln \left( 1-\eta_t \right)}.
\label{eq:fd_1}
\end{equation} Similarly, if the escape times are measured by a non-inertial reference frame comoving with a particle, a fraction of particles $\eta_\tau$ escapes every time $\tau_0$, and therefore the associated fractal dimension can be defined as $d_\tau$.

The behavior is governed by Poisson statistics in the hyperbolic regime. Therefore, the average number of particles that escape follow an exponential decay law. More specifically, the number of particles remaining in the scattering region according to an inertial reference frame at rest in the potential is given by\begin{equation}N(t) = N_0 e ^ {-\sigma t}.\end{equation} We note that this decay is homogeneous in an inertial reference frame, whereas according to an observer describing the decay in a non-inertial reference frame comoving with a particle, we get the decay law \begin{equation}\tilde{N}(\tau) \equiv  N (t (\tau)) = N_0 e^{- \sigma \int_0^\tau \gamma(t (\tau')) d \tau'} ,\end{equation} where we have substituted the equality $ t = \int_{0}^{\tau} \gamma (t (\tau')) d \tau'$ from solving the Eq.~\eqref{eq:dtau}. In other words, for an accelerated observer the decay is still Poissonian, but inhomogeneous. Nevertheless, if we disregard the fluctuations in the Lorentz factor, an homogeneous statistics can be nicely approximated once again, by defining the average constant rate $\bar{\sigma}_{\tau} \equiv \sigma\bar{\gamma}$. We recall that $ \gamma(t) $ is the Lorentz factor along the trajectory of a certain particle, and therefore $ \tilde{N}(\tau)$ here describes the number of particles remaining in the scattering region according to the accelerated frame co-moving with such a particle. This particle must be sufficiently close to the chaotic saddle in order to remain trapped in the well a sufficiently long time so as to render useful statistics, by counting a high number of escaping test bodies.

Now we calculate, without loss of generality, the fraction of particles that escape during an iteration according to this reference frame as \begin{equation} \eta_\tau = \frac{\tilde{N} (\tau_0) - \tilde{N} (\tau_0')}{\tilde{N} (\tau_0)} = \frac{N (t_0) - N (2t_0)}{N (t_0)} = \eta_t,\end{equation} where $\tau_0'$ is the proper time observed by the accelerated body when the clocks at rest in the potential mark $2t_0$. In this manner, we obtain that the fraction of escaping particles is invariant under reference frame transformations, because there exists an unequivocal relation between the times $ t $ and $ \tau $ given by $ \gamma (t) $. From this result we derive that the fractal dimension of the Cantor-like set associated with the escape times function is invariant under coordinate transformations, $ d_t = d_\tau $. This equality holds for every particle clock evolving in the well, as long as it stays long enough. On the other hand, this result is in consonance with the Cantor-like set nature, because its fractal dimension does not depend on how much time an iteration lasts.

In order to compute the fractal dimensions associated with these scattering functions, we make use of the uncertainty dimension algorithm \cite{lau1991,grebogi1983_2} and the shooting method previously described. We launch a particle from the potential minimum with a random shooting angle $\theta_0$ in the interval $[\pi/2,5\pi/6]$ and measure the escape times $t_e(\theta_0)$ and $\tau_e(\theta_0)$, and the exit $e(\theta_0)$ through it escapes. Then, we carry out again the same procedure from a slightly different shooting angle $\theta_0 + \epsilon$, where $\epsilon$ can be considered a small perturbation, and calculate the quantities $t_e(\theta_0+\epsilon)$, $\tau_e(\theta_0+\epsilon)$ and $e(\theta_0+\epsilon)$. We then say that an initial condition $\theta_0$ is \textit{uncertain} in measuring, e.g., the escape time $t_e$, if the difference between the escape times, $|t_e(\theta_0)-t_e(\theta_0+\epsilon)|$, is higher than a given time. This time is usually associated with the integration step $h$ of the numerical method, which is the resolution of an inertial clock. Conveniently, we set this criterion of uncertain initial conditions as $3h/2$, i.e., the half between the step and two times the step of the integrator, for any clock. The reason for it is that the time differences according to a particle clock are the result of a computation by means of Eq.~\eqref{eq:tau_e}. Therefore, an initial condition $\theta_0$ is \textit{uncertain} in measuring the escape time $t_e$ if \begin{equation}
\Delta t_e(\theta_0) = |t_e(\theta_0) - t_e(\theta_0 + \epsilon)| > 3h/2.
\label{eq:un_1}
\end{equation} Similarly, an initial condition $\theta_0$ is uncertain in measuring the escape time $\tau_e$ if \begin{equation}
\Delta \tau_e(\theta_0) = |\tau_e(\theta_0) - \tau_e(\theta_0 + \epsilon)| > 3h/2.
\label{eq:un_2}
\end{equation}
Finally, an initial condition is uncertain with respect to the exit through which the particle escapes if $e(\theta_0) \neq e(\theta_0 + \epsilon)$.

\begin{figure}[b]
    \centering
    \includegraphics[width=0.4\textwidth]{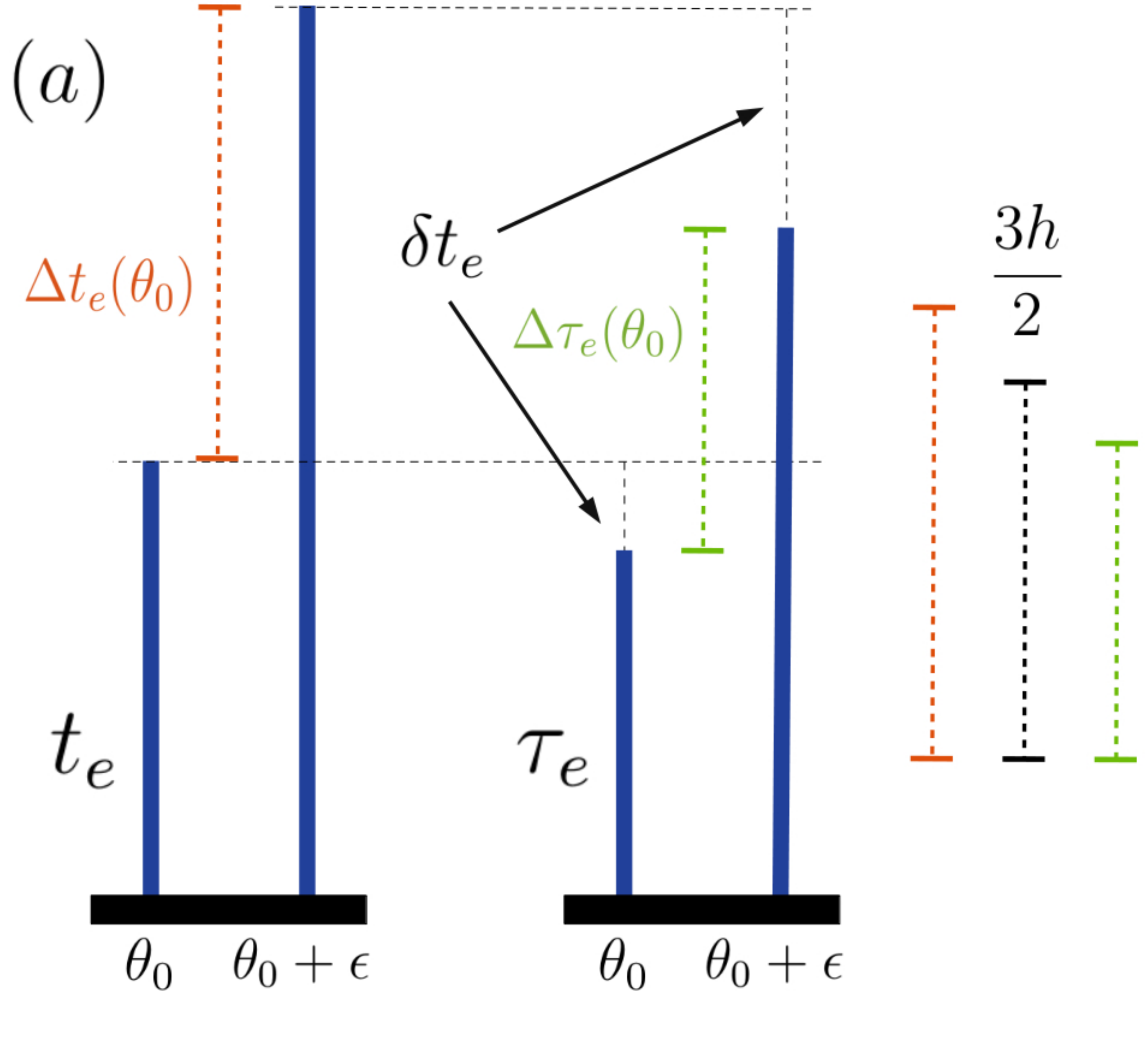}
    \quad
    \includegraphics[width=0.475\textwidth]{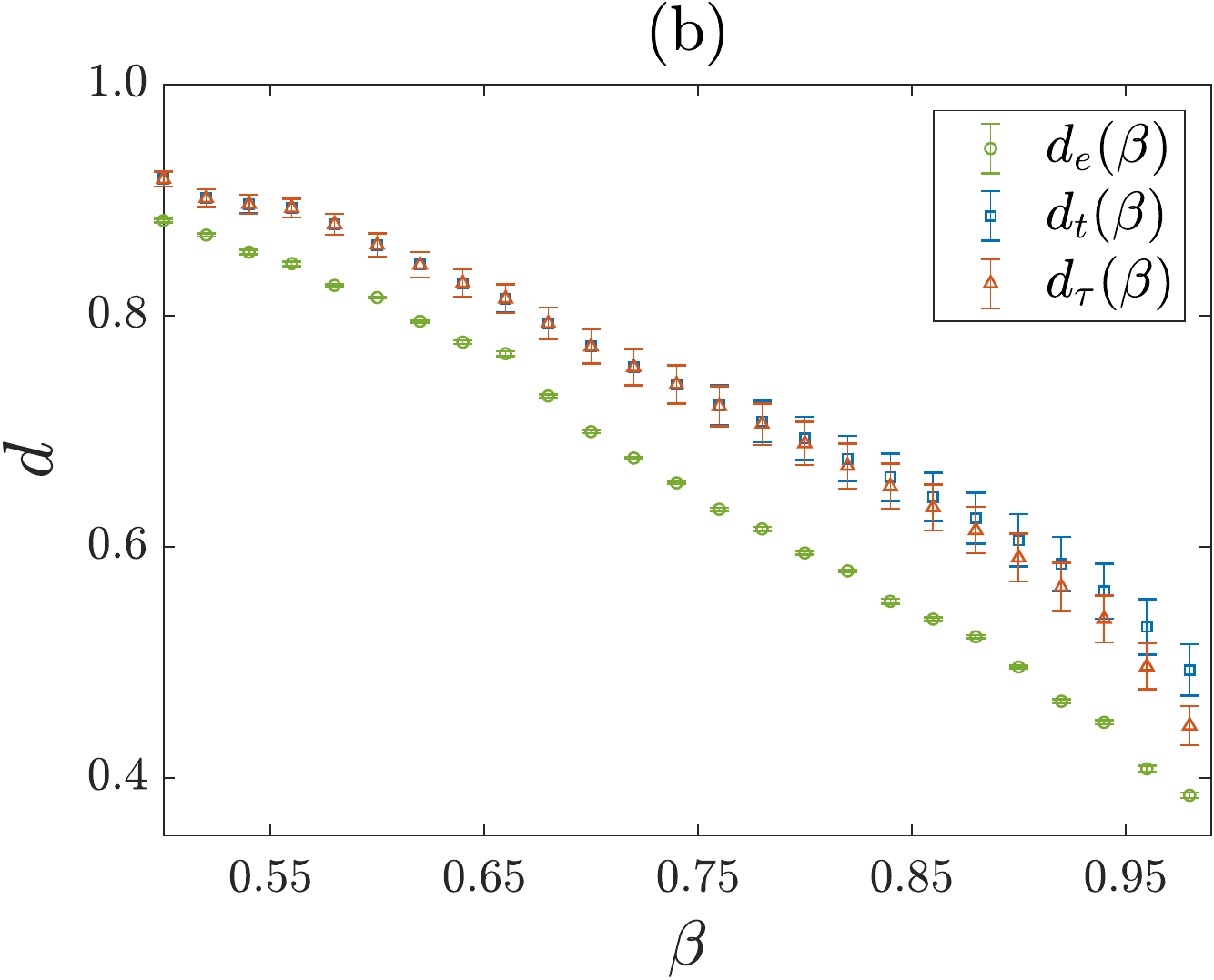}
    \caption{ (a) A scheme to visualize the physical effect of a reference frame modification on the unpredictability of the escape times, where $h = 0.005$. (b) Fractal dimensions according to exits $d_e$ (green), escape time $d_t$ (blue) and escape proper time $d_\tau$ (red) with standard deviations computed by the uncertainty dimension algorithm versus twenty five equally spaced values of $\beta \in [0.5, 0.98]$.}
    \label{fig:sec5}
\end{figure}

We generally expect that the time differences holds $\Delta \tau_e(\theta_0) < \Delta t_e(\theta_0)$, since we have defined previously that $\bar{\tau}_e \equiv t_e/\bar{\gamma}$. Thus, given the same criterion $3h/2$ in both clocks, there will be some uncertain initial conditions $\theta_0$ in the inertial clock $\left( \Delta t_e (\theta_0) > 3h/2 \right)$ that become certain in the particle clock $\left( \Delta \tau_e (\theta_0) < 3h/2 \right)$. We show a scheme in Fig.~\ref{fig:sec5}(a) to clarify this physical effect on the escape times unpredictability. It is easy to see that this effect is caused by the limited resolution of the hypothetical clocks, and becomes more intense for high values of $\beta$ because it is proportional to the Lorentz factor.

The fraction of uncertain initial conditions behaves as \begin{equation} f(\epsilon) \sim \epsilon^{1-d}, \label{eq:un_4}\end{equation} where $d$ is the value of the fractal dimension, which enables us to quantify the unpredictability in foreseeing the particle final dynamical state. In particular, $d = 0$ ($d = 1$) implies minimum (maximum) unpredictability of the system \cite{lau1991}. All the cases in between, $0<d<1$, imply also unpredictability, and the closer to the unity the value of the fractal dimension is, the more unpredictable the system is. According to our scattering functions, it is expected that the values of their fractal dimensions decrease as the value of $\beta$ increase, since these functions become smoother. Taking decimal logarithms in Eq.~\eqref{eq:un_4}, we obtain \begin{equation}
\log_{10} \frac{f(\epsilon)}{\epsilon} \sim -d \log_{10} \epsilon.
\label{eq:logfeps}
\end{equation}
This formula allows us to compute the fractal dimension of the scattering functions from the slope of the linear relation, which obeys a representation $\log_{10} f(\epsilon)/\epsilon$ versus $\log_{10} \epsilon$. We use an adequate range of angular perturbations according to our shooting method and the established criterion of uncertain initial conditions, concretely, $\log_{10} \epsilon \in [-6,-1]$.

The computed fractal dimensions always hold $d_e < d_t, d_\tau$ as shown in Fig.~\ref{fig:sec5}(b). This occurs because it is generally more predictable to determine the exit through which the particle escapes than exactly its escape time when the clocks resolution is small. Therefore, there is a greater number of uncertain conditions concerning escape times than in relation to exits. The former ones are located outside and over the stable manifold, whereas the uncertain conditions regarding exits can only be located on the stable manifold by definition. We obtain computationally $d_t \approx d_\tau$ for almost every value of $\beta$. Nonetheless, the physical effect explained above causes a small difference between the computed fractal dimensions regarding escape times, implying $d_\tau < d_t$ in a very energetic regime.

From a mathematical point of view, if we consider a infinitely small clock resolution, i.e., $h \to 0$, uncertain initial conditions in any clock will be only the ones whose associated escape time differences are also infinitely small. Such uncertain conditions will be located on the stable manifold. In that case, the geometric and observer-independent nature of the fractality caused by the chaotic saddle is reflected into the values of the fractal dimensions. It is expected that in this limit the equality $d_e = d_t = d_\tau$ holds.

This equality extends the very important statement that relativistic \emph{chaos} is coordinate invariant to \emph{transient chaos} as well. The result provided in \cite{motter2003} showing that the signs of the Lyapunov exponents of a chaotic dynamical system are invariant under coordinate transformations can be perfectly extended to transient chaotic dynamics. For this purpose, it is only required to consider a chaotic trajectory on the chaotic saddle, which meets the necessary four conditions described in \cite{motter2003}. Since the sign of the Lyapunov exponents of a trajectory on the chaotic saddle are also invariant, it is therefore evident that the existence of transient chaotic dynamics can not be avoided by considering suitable changes of the reference frame. We believe that this analytical result is at the basis of the results arising from all the numerical explorations performed in the previous sections.

\section{Conclusions} \label{sec:5}

Despite the fact that the H\'{e}non-Heiles system has been widely studied as a paradigmatic open Hamiltonian system, we have added a convenient definition of its scattering region. In this manner, the scattering region can be defined as the part of the physical space where the particle dynamics is bounded, and therefore a particle escapes when its kinetic energy is greater than the kinetic energy value at the potential minimum.

Since relativistic chaos has been demonstrated as coordinate invariant, we have been focused on the special relativistic version of the H\'{e}non-Heiles system to extend this occurrence to transient chaos. We have then analyzed the Lorentz factor effects on the system dynamics, concretely, how the time dilation phenomenon affects the scattering function structure. The exit and the escape time functions exhibit a fractal structure of singularities as a consequence of the presence of the chaotic saddle. Since the origin of the escape time singularities is geometric, the fractality of the escape time function must be independent of the observer. We conclude that the time dilation phenomenon does not affect the typical structure of the singularities of the escape times, and interestingly this phenomenon occurs chaotically.

The escape time function as measured in any clock is closely related to a Cantor-like set of Lebesgue measure zero, since it is a self-similar set in the hyperbolic regime. This feature allows us to demonstrate that the fractal dimension of the escape time function is relativistic invariant. The key point of the demonstration is that, knowing the evolution of the Lorentz factor, there exists an unequivocal relation between the transformed times. In order to verify this result computationally, we have used the uncertainty dimension algorithm. Furthermore, we have pointed out that the system is more likely to be predictable in a reference frame comoving with the particle if a limited clock resolution is considered, even though from a mathematical point of view the predictability of the system is independent of the reference frame.

The main conclusion of the present work is that transient chaos is coordinate invariant from a theoretical point of view. This statement extends the universality of occurrence of chaos and fractals under coordinate transformations to the realm of transient chaotic phenomena as well.

\section*{ACKNOWLEDGMENTS}

We acknowledge interesting discussions with Prof. Hans C. Ohanian. This work was supported by the Spanish State Research Agency (AEI) and the European Regional Development Fund (ERDF) under Project No.~FIS2016-76883-P.


\begin{thebibliography}{90}

        \bibitem{seoane2013}
    J. M. Seoane and M. A. F. Sanjuán, Rep. Prog. Phys. \textbf{76}, 016001 (2013).

        \bibitem{lin2013}
    Y.-D. Lin, A. M. Barr, L. E. Reichl, and C. Jung, Phys. Rev. E \textbf{87}, 012917 (2013).

        \bibitem{daitche2014}
    A. Daitche and T. T\'{e}l, New J. Phys. \textbf{16}, 073008 (2014).

        \bibitem{zotos2017}
    E. E. Zotos and C. Jung, Mon. Notices Royal Astron. Soc. \textbf{465}, 525–546 (2017)

        \bibitem{navarro2019}
    J. F. Navarro, Sci. Rep. \textbf{9}, 13174 (2019).

        \bibitem{lai2010}
    Y.-C. Lai and T. T\'{e}l, \textit{Transient Chaos: Complex Dynamics on Finite-Time Scales}, Springer, New York (2010).

        \bibitem{grebogi1983}
    C. Grebogi, E. Ott, and J. A. Yorke, Physica D \textbf{7}, 181 (1983).

        \bibitem{aguirre2009}
    J. Aguirre, R. L. Viana, and M. A. F. Sanju\'{a}n, Rev. Mod. Phys. \textbf{81}, 333 (2009).

        \bibitem{ott1993}
    E. Ott, \textit{Chaos in Dynamical Systems} (Cambridge University Press, New York, NY, 1993).

        \bibitem{tel2015}
    T. T\'{e}l, Chaos \textbf{25}, 097619 (2015).

        \bibitem{hobill1994}
    D. Hobill, A. Burd, and A. Coley, \textit{Deterministic Chaos in General Relativity} (Plenum, New York, 1994).

        \bibitem{motter2003}
    A. E. Motter, Phys. Rev. Lett. \textbf{91}, 231101 (2003).

        \bibitem{motter2009}
    A. E. Motter and A. Saa, Phys. Rev. Lett. \textbf{102}, 184101 (2009).

        \bibitem{vallejo2003}
    J. C. Vallejo, J. Aguirre, and M. A. F. Sanjuán, Phys. Lett. A \textbf{311}, 26–38 (2003).

        \bibitem{aguirre2001}
    J. Aguirre, J. C. Vallejo, and M. A. F. Sanju\'{a}n, Phys. Rev. E \textbf{64}, 066208 (2001).

        \bibitem{motter2001}
    A. E. Motter and P. S. Letelier, Phys. Lett. A \textbf{285}, 127–131 (2001).

        \bibitem{barrow1982}
    J. D. Barrow, Gen. Relat. Gravit. \textbf{14}, 523-530 (1982).

        \bibitem{chernikov1989}
    A. A. Chernikov, T. T\'{e}l, G. Vattay, and G. M. Zaslavsky, Phys. Rev. A \textbf{40}, 4072 (1989).

        \bibitem{ni2012}
    X. Ni, L. Huang, Y.-C Lai, and L. M. Pecora, EPL \textbf{98}, 50007 (2012).

        \bibitem{bernal2017}
    J. D. Bernal, J. M. Seoane, and M. A. F. Sanju\'{a}n, Phys. Rev. E \textbf{95}, 032205 (2017).

        \bibitem{bernal2018}
    J. D. Bernal, J. M. Seoane, and M. A. F. Sanju\'{a}n, Phys. Rev. E \textbf{97}, 042214 (2018).

        \bibitem{henon1964}
    M. H\'{e}non and C. Heiles, Astron. J. \textbf{69}, 73 (1964).

        \bibitem{contopoulos1990}
    G. Contopoulos, Astron. Astrophys. \textbf{231}, 41 (1990).

        \bibitem{sideris2006}
    I. V. Sideris, Phys. Rev. E \textbf{73}, 066217 (2006).

        \bibitem{ohanian2001}
    H. O. Ohanian,  \textit{Special Relativity: A Modern Introduction}, Physics Curriculum $\&$ Instruction, Inc, First Edition (2001).

        \bibitem{lan2011}
    B. L. Lan and F. Borondo, Phys. Rev. E \textbf{83}, 036201 (2011).

        \bibitem{chanda2018}
    S. Chanda and P. Guha, Int. J. Geom. Methods Mod. Phys. \textbf{15}, 1850062 (2018).

        \bibitem{kovacs2011}
    T. Kov\'{a}cs, Gy. Bene, and T. T\'{e}l, Mon. Not. R. Astron. Soc. \textbf{414}, 2275–2281 (2011).

        \bibitem{calura1997}
    M. Calura, P. Fortini, and E. Montanari, Phys. Rev. D \textbf{56}, 4782 (1997).

        \bibitem{press1992}
    W.H. Press, B.P. Flannery, S.A. Teukolsky, and W.T. Vetterling, \textit{Numerical Recipes in C: The Art of Scientific Computing}, Cambridge Univ. Press (1992).

        \bibitem{barton1999}
    G. Barton, \textit{Introduction to the Relativity Principle: Particles and Plane Waves}, John Wiley \& Sons (1999).

        \bibitem{jeffreys1988}
    H. Jeffreys and B. S. Jeffreys, \textit{Methods of Mathematical Physics}, 3rd ed., Cambridge University Press (1988).

        \bibitem{seoane2007}
    J. M. Seoane, M. A. F. Sanju\'{a}n, and Y.-C. Lai,  Phys. Rev. E \textbf{76}, 016208 (2007).

        \bibitem{lau1991}
    Y.-T. Lau, J. M. Finn, and E. Ott, Phys. Rev. Lett. \textbf{66}, 978 (1991).

        \bibitem{grebogi1983_2}
    C. Grebogi, S. W. McDonald, E. Ott, and J. A. Yorke, Phys. Lett. A \textbf{99}, 415 (1983).

\end{thebibliography}
\end{document}